\documentclass[twocolumn,showpacs,prb]{revtex4}
\usepackage{graphicx}
\usepackage{latexsym}

\begin{document}

\title{AlAs 2D electrons in an antidot lattice: Electron pinball with elliptical Fermi contours}
\author{O. Gunawan, E. P. De Poortere and M. Shayegan}
\affiliation{Department of Electrical Engineering, Princeton University, Princeton, NJ 08544}
\date{\today}

\begin{abstract}
We report ballistic transport measurements in a two-dimensional electron system confined to an
AlAs quantum well and patterned with square antidot lattices of period $a = $0.6, 0.8, 1.0 and 1.5
$\mu$m. In this system two in-plane conduction-band valleys with elliptical Fermi contours are
occupied. The low-field magneto-resistance traces exhibit peaks corresponding to the
commensurability of the cyclotron orbits and the antidot lattice. From the dependence of the
position of the peak associated with the smallest commensurate orbit on electron density and $a$,
we deduce the ratio of the longitudinal and transverse effective masses $m_l/m_t=5.2\pm 0.4$, a
fundamental parameter for the anisotropic conduction bands in AlAs.
\end{abstract}

\pacs{72.20.-i, 73.23.Ad, 75.47.Jn}

\maketitle

\begin{figure}
\includegraphics[width=1\linewidth]{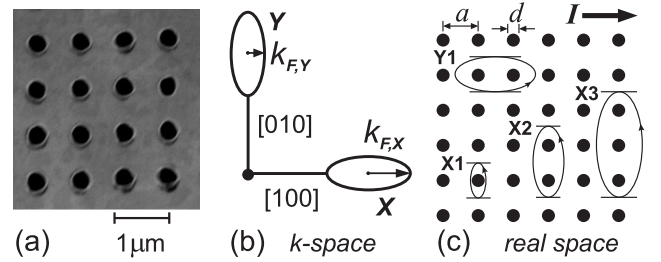} \caption{
(a) Micrograph of the AD lattice region with period $a$=0.8 $\mu$m. (b) The Fermi contours of AlAs
in-plane valleys $X$ and $Y$ in $k$-space. The Fermi wave vectors $k_{F,X}$ and $k_{F,Y}$ are
indicated. (c) The first four commensurate orbits for the $X$ and $Y$ valleys that give rise to
peaks in magneto-resistance for current along the $x$-direction; these orbits have diameters in
the $y$-direction that are equal to a multiple integer of the AD lattice period (see text).}
\label{FigADPic}
\end{figure}

The crystal structure and lattice constant of AlAs are closely matched to that of GaAs and, as a
result, AlAs is widely used as a barrier material in AlAs/GaAs hetero-structures. However, AlAs
differs greatly from GaAs in its band-structure. In contrast to GaAs where electrons occupy a
single, isotropic, conduction-band minimum at the center of the Brillouin zone, electrons in AlAs
populate anisotropic conduction-band minima (valleys) centered at the X-points of the Brillouin
zone. The constant energy surface in (bulk) AlAs consists of six, anisotropic, half-ellipsoids
(three full-ellipsoids). It is also possible to confine electrons to a modulation-doped AlAs
quantum well, thereby forming a high-mobility two-dimensional electron system (2DES).\cite{AlAsQW}
If the quantum well is wider than about 5 nm, the electrons occupy only the two conduction-band
valleys with their major axes in the plane, along the [100] and [010] crystal axes. We refer to
these two valleys as $X$ and $Y$, respectively [Fig.~\ref{FigADPic}(b)]. Here we report ballistic
transport measurements in a square anti-dot (AD) array \cite{WeissPRL91, LorkePRB91,
FleischmannPRL92} patterned in a high-mobility AlAs 2DES. The data provide a direct measure of the
anisotropy of the Fermi contour, or equivalently, the ratio of the longitudinal and transverse
electron effective masses, a fundamental parameter of the AlAs conduction-band. Recently it has
been shown that this anisotropy can be exploited to realize a simple "valley-filter" device using
a quantum point contact structure.\cite{GunawanPRB06} Such a device may play an important role in
"valleytronics" or valley-based electronics applications,\cite{Rycerz06} or for quantum
computation where the valley state of an electron might be utilized as a qubit.\cite{GunawanPRL06}

\begin{figure}
\includegraphics[width=1\linewidth]{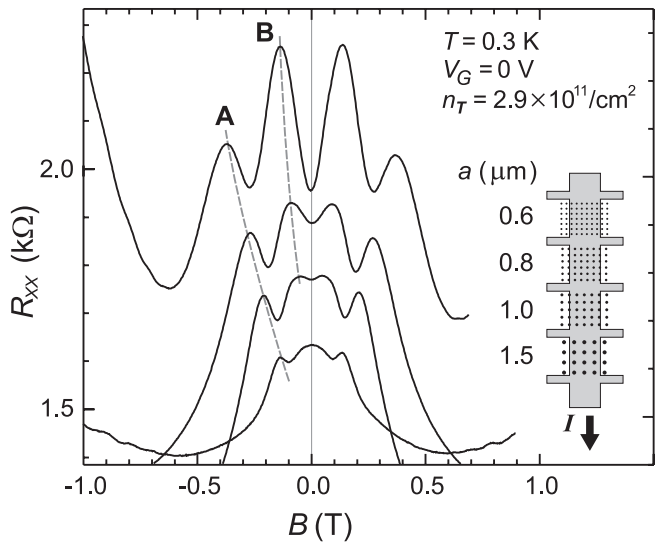} \caption{
Low-field MR  traces for all four AD regions with periods (from top to bottom) equal to $a$ = 0.6,
0.8, 1.0, and 1.5 $\mu$m. Traces are offset vertically for clarity, except for the bottom one. The
Hall bar with the different AD regions is schematically shown on the right.} \label{FigMRall}
\end{figure}

We performed experiments on 2DESs confined to a high-quality, modulation-doped, 15 nm-wide, AlAs
quantum well grown by molecular beam epitaxy on a (001) GaAs substrate.\cite{AlAsQW} The AlAs
quantum well is flanked by undoped layers of Al$_{0.4}$Ga$_{0.6}$As on each side, and the doping
is via a delta-layer of Si on the surface side at a distance of 100 nm from the well and 32 nm
from the surface. We patterned a Hall bar sample, with the current direction along [100], using
standard optical photolithography. We then deposited a layer of polymethyl methacrylate (PMMA) and
patterned the AD arrays using electron beam lithography. The PMMA layer served as a resist for a
subsequent dry etching process used to define the AD holes. We used an electron cyclotron
resonance etching system with an Ar/Cl$_2$ plasma,\cite{ChenJVSTB00} at an etch rate of $\simeq$55
nm/min, to obtain small feature sizes without a degradation of the 2DES quality. The AD pattern
was etched to a depth of $\simeq$80 nm, thus stripping the dopant layer and depleting the
electrons in the AD regions. The micrograph of a section of one of our AD arrays is shown in
Fig.~\ref{FigADPic}(a). Each AD array is a square lattice and covers a 20 $\mu$m $\times$ 30
$\mu$m  area. There are four regions of AD lattice with different periods: $a$ = 0.6, 0.8, 1.0,
and 1.5 $\mu$m as schematically shown in the inset of Fig.~\ref{FigMRall}. The aspect ratio $d/a$
of each AD cell is $\sim$1:3, where $d$ is the AD diameter. Finally, we deposited a front gate,
covering the entire surface of the active regions of the sample to control the 2DES density.
Following an initial back-gate biasing and brief illumination,\cite{PoorterePRB03} we used the
front gate to tune the total density ($n_T$) from $2$ to 5$\times$$10^{11}$/cm$^2$. The  density
was determined from both Shubnikov-de Haas oscillations and Hall coefficient measurements that
agree with each other. From the measurements on an unpatterned Hall bar region in a different
sample but from the same wafer, we obtain a mobility of $\sim$10 m$^2$/Vs at a typical density of
3$\times$$10^{11}$/cm$^2$ and $T$=0.3 K. This gives a typical mean-free-path of $\sim$1 $\mu$m.

\begin{figure}
\includegraphics[width=1\linewidth]{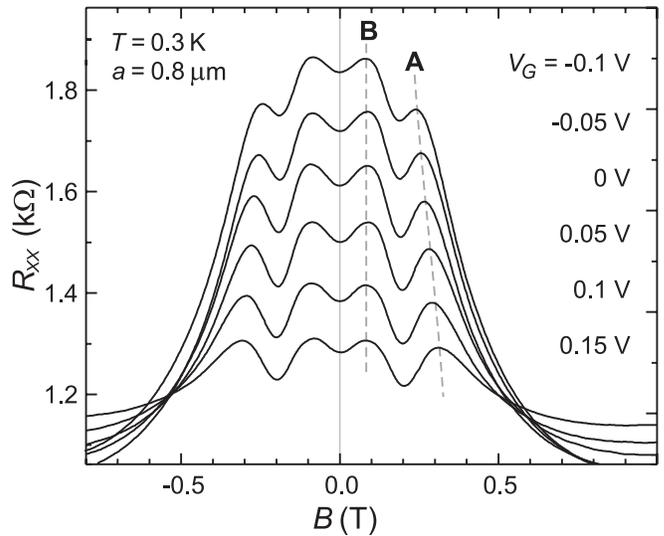} \caption{
Low-field MR traces for the AD region with period $a=0.8$ $\mu$m for $V_G$ = -0.1 V to 0.15 V
(from top to bottom), corresponding to a linear variation of the density $n_T$ from 2.27 to
3.53$\times$$10^{11}$/cm$^2$. Traces are offset for clarity, except for the bottom one. }
\label{FigMRVFG}
\end{figure}

Figures~\ref{FigMRall} and \ref{FigMRVFG} summarize our main experimental results.
Figure~\ref{FigMRall} shows the low-field magneto-resistance (MR) traces, measured as a function
of perpendicular magnetic field ($B$), for all the AD regions. We observe two peaks, A and B,
which are symmetric with respect to $B$$=$$0$. Peak A, whose position is higher in field than peak
B, is seen in the traces from all the AD regions. Peak B, on the other hand, is not observed in
the $a$=1.5 $\mu$m trace. In general, we observe that, as the AD lattice period becomes smaller,
the positions of both peaks A and B shift to higher field values (as indicated by the dashed
lines). Figure~\ref{FigMRVFG} captures the gate-voltage ($V_G$) dependence of the MR traces for
the $a$=0.8 $\mu$m AD region. As we increase $V_G$ to increase the 2DES density, peak A shifts to
higher field values while peak B does not appear to shift. We have made similar observations in
the other AD regions as $V_G$ is varied.

\begin{figure}
\includegraphics[width=1\linewidth]{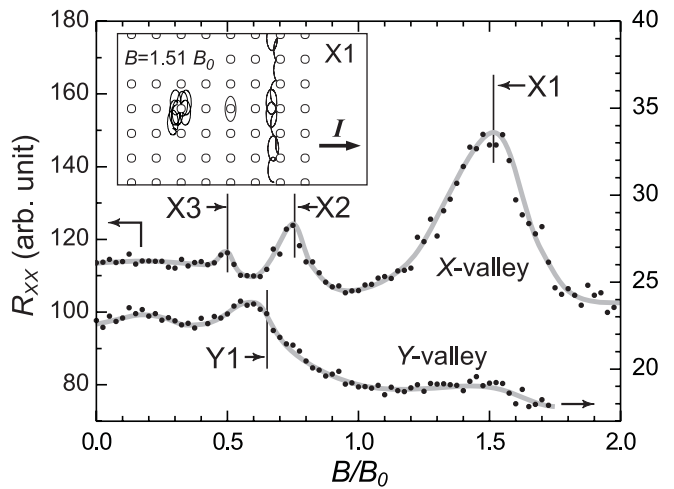}
\caption{MR obtained from numerical simulations (smooth curves are guides to the eye). Vertical
lines indicate the expected positions of the peaks for orbits X1, X2, X3 and Y1, based on
Fig.~\ref{FigADPic}(c). We assumed equal densities for the two valleys and a current along the
$x$-direction. $B_0$ is the magnetic field of the first commensurate orbit if the Fermi contour
were circular. Inset: Simulation snapshot showing various types of trajectories: chaotic, pinned
and skipping orbits, for $X$-valley electrons when X1 peak occurs ($B=\sqrt[4]{m_l/m_t}\,B_0$).}
\label{FigADSim}
\end{figure}

In order to analyze the data of Figs.~\ref{FigMRall} and \ref{FigMRVFG}, we first briefly review
what is known about ballistic transport in AD arrays for \textit{GaAs} 2DESs where the Fermi
contour is \textit{isotropic}. Low-field MR traces for such systems typically exhibit
commensurability peaks at magnetic fields where the classical cyclotron orbit fits around a group
of ADs.\cite{WeissPRL91, LorkePRB91, FleischmannPRL92} Although there are subtleties associated
with the exact shape of the AD potential and also the possibility of chaotic orbits that bounce
from one AD boundary to another, the peak observed at the highest magnetic field corresponds to
the shortest period that fits around the smallest number of ADs. For an isotropic Fermi contour,
this corresponds to a circular orbit, with a diameter equal to the AD period, encircling a single
AD. There have also been studies of ballistic transport in 2DESs with isotropic orbits in an
\textit{anisotropic} (rectangular) AD lattice. \cite{ADEquivalence} Experimental results,
\cite{TsukagoshiPRB95} followed by theoretical analysis,\cite{LuPRB96} indicated that the
commensurability peaks are observed only when the orbit diameter matches an integer multiple of AD
lattice period \textit{along the direction perpendicular to the current}.\cite{GridFocusing} Based
on the above considerations, we can predict the first four (smallest) commensurate orbits of the
$X$ and $Y$ valleys that may give rise to MR peaks in our system; these are shown in
Fig.~\ref{FigADPic}(c).

To evince this conjecture, we performed a kinematic, numerical simulation for our system, a 2DES
with elliptical Fermi contours in an isotropic AD lattice in the presence of a perpendicular
magnetic field $B$. The simulation details are similar to those described in
Ref.~\onlinecite{NagaoJPSJ95}. The AD boundaries are represented as hard-wall potentials where the
electrons are scattered elastically upon collision. A set of typical electron trajectories are
shown in the inset of Fig.~\ref{FigADSim}. Using the Kubo formula\cite{KuboJPSJ57, NagaoJPSJ95} we
calculated the conductivity tensor and then obtained the resistance through its inversion. We
performed the simulations for two separate cases, one for $X$ and one for $Y$-valley electrons and
present the results in Fig.~\ref{FigADSim}. The resulting MR traces indeed show peaks at or near
the expected values, namely orbits X1, X2, X3 for the $X$-valley and Y1 for the $Y$-valley
[Fig.~\ref{FigADPic}(c)].

Now we focus our attention back on the experimental results. We associate peak A in our data of
Figs.~\ref{FigMRall} and \ref{FigMRVFG} with the shortest orbit X1 in Fig.~\ref{FigADPic}(c). From
the field position of this peak, and if we assume that the electron density of the $X$ valley is
half the total density, we can directly obtain a value for the anisotropy (ratio of the major to
minor axes diameters) of the elliptical orbits in our system, thus obtaining the effective mass
ratio $m_l/m_t$.\cite{OrbitAnisotropy} However, there is a finite imbalance between the $X$ and
$Y$ valley densities in our sample. Such imbalances can occur because of anisotropic strain in the
plane of the sample and are very often present in AlAs 2DESs.\cite{AlAsQW, GunawanPRL04}
Therefore, we present here an analysis to determine the $m_l/m_t$ ratio independent of the density
imbalance.

Consider a primary, commensurate orbit whose diameter in the direction perpendicular to the
current is equal to the AD lattice period [orbits X1 and Y1 in Fig.~\ref{FigADPic}(c)]. These
would give rise to MR peaks at fields $B_P =2\hbar k_F/e a$ where $k_F$ is the Fermi wavevector
$along$ the current direction. For the $X$ and $Y$ valleys, these wavevectors are $k_{F,X}$ and
$k_{F,Y}$, respectively, as shown in Fig.~\ref{FigADPic}(b). For an elliptical Fermi contour, they
are related to the densities of the $X$ and $Y$ valleys, $n_X$ and $n_Y$, via the following
relations:
\begin{eqnarray}
  k^2_{F,X}\!=\!2\pi n_{X} \sqrt{m_l/m_t},\qquad k^2_{F,Y}\!=\!2\pi n_{Y} \sqrt{m_t/m_l}\,.\label{EqkF}
\end{eqnarray}
Note that the total density $n_T$$=$$n_X$$+$$n_Y$ and the valley imbalance $\Delta
n$$=$$n_X$$-$$n_Y$. We can obtain $n_T$ from the Shubnikov-de Haas oscillations of the MR at high
magnetic fields or from a measurement of the Hall coefficient. Now consider orbit X1 as shown in
Fig.~\ref{FigADPic}(c). Its associated MR peak position $B_{P,X1}$ is given as:
\begin{equation}
B_{P,X1}^{\,2}=\frac{h^2}{\pi e^2 a^2} \sqrt{m_l\over m_t}(n_T+\Delta n). \label{EqBp2}
\end{equation}
We use this expression to analyze our data.

\begin{figure}
\includegraphics[width=1\linewidth]{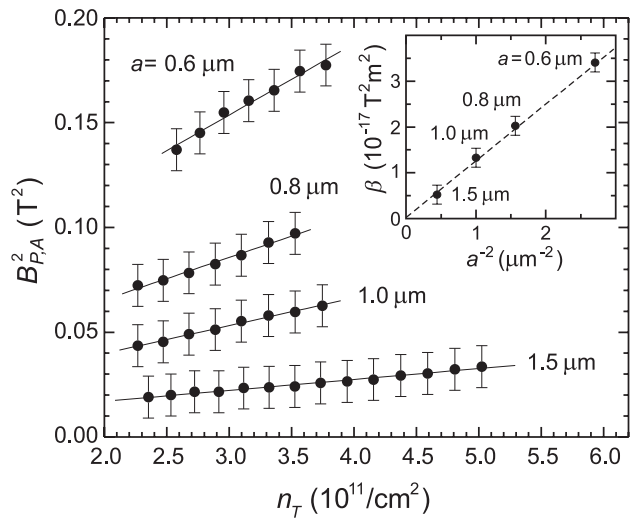}
\caption{Summary of the density dependence of $B_{P,A}^{\,2}$ for all four AD regions; $B_{P,A}$
is the position of peak A observed in MR traces. The straight lines are linear fits using
Eq.~\ref{EqBp2}. The error bars reflect the uncertainty in the peak positions which were
determined by subtracting second-order polynomial backgrounds from the MR traces. Inset: Slope
($\beta$) of the $B_{P,A}^{\,2}$ vs $n_T$ lines of the main figure are plotted as a function of
$a^{-2}$. The dashed-line is a linear fit to the data.} \label{FigBp2vsN}
\end{figure}

We assign peak A in our data to orbit X1 and plot the square of its field position $B_{P,A}^{\,2}$
as a function of $n_T$ in Fig.~\ref{FigBp2vsN}. It is clear that for all four AD lattice regions,
$B_{P,A}^{\,2}$ varies linearly with $n_T$ as expected from Eq.~\ref{EqBp2}. Moreover, we obtain
the slopes ($\beta$) and the intercepts of the lines in Fig.~\ref{FigBp2vsN} by performing a
least-squares fit of each data set. Note that according to Eq.~\ref{EqBp2}, $\beta=h^2
\sqrt{m_l/m_t}/\pi e^2a^2$, and the intercept is equal to $\beta \Delta n$.\cite{DeltaN} Finally,
we plot $\beta$ as a function of $a^{-2}$ in the inset of Fig.~\ref{FigBp2vsN}. This figure shows
that, consistent with the prediction of Eq.~\ref{EqBp2}, $\beta$ indeed depends linearly on
$a^{-2}$ and the line has a zero intercept.

From the slope, $\Delta \beta/\Delta (a^{-2})$, of the line in Fig.~\ref{FigBp2vsN} inset, we can
deduce the effective mass anisotropy ratio:
\begin{equation}
m_l/m_t=\pi^2 e^4/h^4 \left[\Delta \beta/\Delta (a^{-2})\right]^2. \label{Eqmlmt}
\end{equation}
Note that this mass anisotropy ratio is related to the slope of the line in Fig.~\ref{FigBp2vsN}
inset by a pre-factor containing only physical constants ($\pi^2 e^4/h^4$). Our data analysis and
determination of this ratio are therefore insensitive to parameters such as density imbalance
between the two valleys. From data of Fig.~\ref{FigBp2vsN} we obtain $m_l/m_t=5.2\pm0.4$, in very
good agreement with the ratio $m_l/m_t=5.2\pm0.5$ determined from recent ballistic transport
measurements in AlAs 2DESs subjected to one-dimensional, periodic potential
modulations.\cite{GunawanPRL04} Given the cyclotron mass in AlAs, $m^{\ast}$=$\sqrt{m_l m_t}=0.46
\,m_0$,\cite{LayAPL93} we deduce $m_l=(1.05$$\pm$$0.1)m_0$ and $m_t=(0.20$$\pm$$0.02)m_0$, where
$m_0$ is the free electron mass, in good agreement with the results of the majority of theoretical
and experimental determinations of the effective mass in AlAs.\cite{GunawanPRL04}

Several other features of the data presented here are noteworthy. From the intercepts of the
linear fits in Fig.~\ref{FigBp2vsN} we can determine the valley density imbalance for each AD
lattice. Such analysis gives $\Delta n$ = 1.5, 1.2, 1.0, and 1.3 $\times$10$^{11}$ /cm$^2$ ($\pm
0.2$$\times$10$^{11}$ /cm$^2$) for the AD regions with $a$ = 0.6, 0.8, 1.0, and 1.5 $\mu$m,
respectively. Such a variation of valley imbalance for different AD regions may come from
non-uniform residual strain across the sample.  Note that, because of the close proximity of the
different AD lattice regions, we expect this variation to be small, consistent with the $\Delta n$
values deduced from the above analysis.

\begin{figure}
\includegraphics[width=1\linewidth]{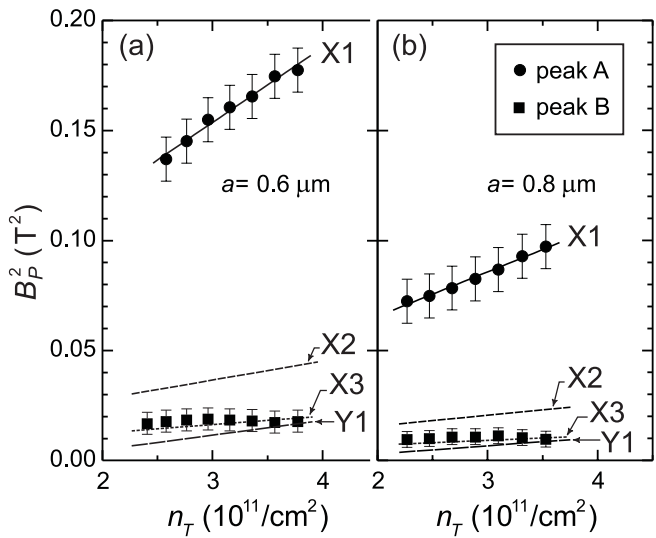} \caption{
Summary of the density dependence of $B_{P}^{\,2}$ of peak A and B for AD region: (a) $a$ = 0.6
$\mu$m and (b) $a$ = 0.8 $\mu$m. The lines marked X1 are linear fits to peak A using
Eq.~\ref{EqBp2}. Lines X2, X3 and Y1 are the predicted peak positions calculated using equations
similar to Eq.~\ref{EqBp2} (see text).}\label{FigBp2XY}
\end{figure}

As for peak B, it is tempting to associate it with orbits X2 or Y1 in Fig.~\ref{FigADPic}(c). This
is qualitatively consistent with the data of Fig.~\ref{FigMRall} which indicate that peak B moves
to smaller values of magnetic field as the period of the AD lattice is made larger. Moreover, peak
B becomes weaker with increasing AD lattice period and disappears for the largest period $a=1.5$
$\mu$m. This is also consistent with the larger size of the X2 and Y1 orbits (compared to the X1
orbit), and the fact that for $a=1.5$ $\mu$m, the lengths of these orbits become large compared to
the electron mean-free-path. Quantitatively, using the values of $m_l/m_t$ and $\Delta n$ obtained
above, we can modify Eq.~\ref{EqBp2} and determine the expected peak positions associated with the
X2, X3 and Y1 orbits.\cite{ModEqBp2} As illustrated in Fig.~\ref{FigBp2XY}, we find that the
predicted peaks for X2, X3 and Y1 orbits are close to each other in field and approximately
straddle the observed positions of peak B in Fig.~\ref{FigMRall}.\cite{X2X3Y1} It is possible then
that peak B may originate from a superposition of X2, X3 and Y1 peaks that cannot be resolved in
our experiment. We cannot rule out, however, that peak B may be strongly influenced by non-linear
orbit resonances in the system. Such resonances are known to occur for orbits with long
trajectories in the presence of a smooth AD potential. \cite{FleischmannPRL92}

In conclusion, we performed ballistic transport experiments in AD lattices imposed on an AlAs 2DES
where the electrons occupy two valleys with anisotropic Fermi contours. The low-field MR traces
exhibit two sets of peaks. From the analysis of the positions of the peak associated with a
commensurate orbit with the shortest trajectory [X1 orbit in Fig.~\ref{FigADPic}(c)], we deduced
the effective mass anisotropy ratio $m_l/m_t$, a fundamental parameter of the AlAs conduction-band
dispersion that cannot be directly measured from other transport experiments.

Our work was supported by the ARO, NSF, and the Alexander von Humboldt Foundation. We thank I.\
Trofimov and G.\ Sabouret for assistance with the electron cyclotron resonance etcher, and Y.\ P.\
Shkolnikov, K.\ Vakili, and R.\ Winkler for illuminating discussions.


\begin{thebibliography}{21}
\bibitem[{ADE()}]{ADEquivalence}
\bibinfo{note}{The problem of ballistic transport for a 2DES with a circular
  Fermi contour in a rectangular AD lattice is equivalent to transport in a
  2DES with an elliptical Fermi contour in a square AD lattice.}
\bibitem[{Orb()}]{OrbitAnisotropy}
\bibinfo{note}{The ratio of the major to minor orbit diameter is
  $\sqrt{m_l/m_t}$.}
\bibitem[{Del()}]{DeltaN}
\bibinfo{note}{We assume $\Delta n$ is fixed in the density range of interest.}
\bibitem[{X2X()}]{X2X3Y1}
\bibinfo{note}{In Fig.~\ref{FigBp2XY} the expected $B_P$ for Y1 orbits are
  lower than $B_P$ for X3 because $n_Y < n_X$ in our experiments.}
\end{thebibliography}
\end{document}